\documentclass{sigchi}

\usepackage{xcolor}

\CopyrightYear{2018} 
\setcopyright{acmlicensed}
\doi{http://dx.doi.org/10.475/123_4}
\isbn{123-4567-24-567/08/06}
\acmPrice{\$15.00}

\usepackage{balance}       
\usepackage{graphics}      
\usepackage[T1]{fontenc}   
\usepackage{txfonts}
\usepackage{mathptmx}
\usepackage[pdflang={en-US},pdftex]{hyperref}
\usepackage{color}
\usepackage{booktabs}
\usepackage{textcomp}

\usepackage{enumitem}
\usepackage{microtype}        
\usepackage{ccicons}          
\usepackage{makecell}

\usepackage{todonotes}

\newcommand{\Comment}[1]{}

\newcommand{\etal}{\mbox{\it et al.}}

\newcommand{\eg}{\mbox{\it e.g.}}
\newcommand{\ie}{\mbox{\it i.e.}}

\newcommand{\unscsys}{\mbox{Cicero}}
\newcommand{\sys}{\mbox{\sc \unscsys}}
\newcommand{\implement}{\mbox{\sc \sys -Sync}}

\def\plaintitle{\unscsys: Multi-Turn, Contextual Argumentation for Accurate Crowdsourcing}

\def\emptyauthor{}
\def\plainkeywords{Crowdsourcing; argumentation; dialog}

\makeatletter
\def\url@leostyle{%
  \@ifundefined{selectfont}{
    \def\UrlFont{\sf}
  }{
    \def\UrlFont{\small\bf\ttfamily}
  }}
\makeatother
\def\pprw{8.5in}
\def\pprh{11in}

\definecolor{linkColor}{RGB}{6,125,233}
\hypersetup{%
  pdftitle={\plaintitle},
  pdfauthor={\emptyauthor},
  pdfkeywords={\plainkeywords},
  pdfdisplaydoctitle=true, 
  bookmarksnumbered,
  pdfstartview={FitH},
  colorlinks,
  citecolor=black,
  filecolor=black,
  linkcolor=black,
  urlcolor=linkColor,
  breaklinks=true,
  hypertexnames=false
}

\begin{document}

\title{\plaintitle}

\numberofauthors{1}
\author{
  \alignauthor Quanze Chen\textsuperscript{1}, Jonathan Bragg\textsuperscript{1}, Lydia B. Chilton\textsuperscript{2}, Daniel S. Weld\textsuperscript{1}\\
  \affaddr{\textsuperscript{1}University of Washington, Seattle, WA, USA} \\
  \affaddr{\textsuperscript{2}Columbia University, New York, NY, USA} \\
  \email{\{cqz, jbragg, weld\}@cs.washington.edu, chilton@cs.columbia.edu}
}

\maketitle

\begin{abstract}
Traditional approaches for ensuring high quality crowdwork have failed to achieve high-accuracy on difficult problems. Aggregating redundant answers often fails on the hardest problems when the majority is confused. Argumentation has been shown to be effective in mitigating these drawbacks. However, existing argumentation systems only support limited interactions and show workers general justifications, not context-specific arguments targeted to their reasoning.

This paper presents \sys, a new workflow that improves crowd accuracy on difficult tasks by engaging workers in multi-turn, contextual discussions through real-time, synchronous argumentation. Our experiments show that compared to previous argumentation systems (\eg, {\sc Microtalk}~\cite{Drapeau2016MicroTalkUA}) which only improve the average individual worker accuracy by 6.8 percentage points on the Relation Extraction domain, our workflow achieves 16.7 percentage point improvement.  Furthermore, previous argumentation approaches don't apply to tasks with many possible answers; in contrast, \sys\ works well in these cases, raising  accuracy from 66.7\% to 98.8\%\ on the Codenames domain. 

\end{abstract}

\category{H.5.m.}{Information Interfaces and Presentation
  (e.g. HCI)}{Group and Organization Interfaces} 




\section{Introduction}

Crowdsourcing has been used for a wide variety of tasks, from image labeling to language transcription and translation. Many complex jobs can be decomposed into small micro-tasks
~\cite{little-hcomp09,bernstein2010soylent,noronha2011platemate,chilton-chi13}. After such decomposition, the primary challenge becomes ensuring that independent individual judgments result in accurate global answers. Approaches ranging aggregation via majority vote~\cite{snow-emnlp08} to programmatic filtering via gold-standard questions~\cite{oleson-hcomp2011} have all been created to achieve this goal. Further improvements have led to more intelligent aggregation such as expectation maximization (EM)~\cite{dawid-as79,whitehill-nips09,welinder-nips10}. However, EM may still fall short, especially on hard problems where individual judgments are unreliable. Indeed, some researchers have concluded that crowdsourcing is incapable of achieving perfect accuracy~\cite{demartini-www12}.

\begin{figure}
 \centering
 \includegraphics[width=0.84\columnwidth]{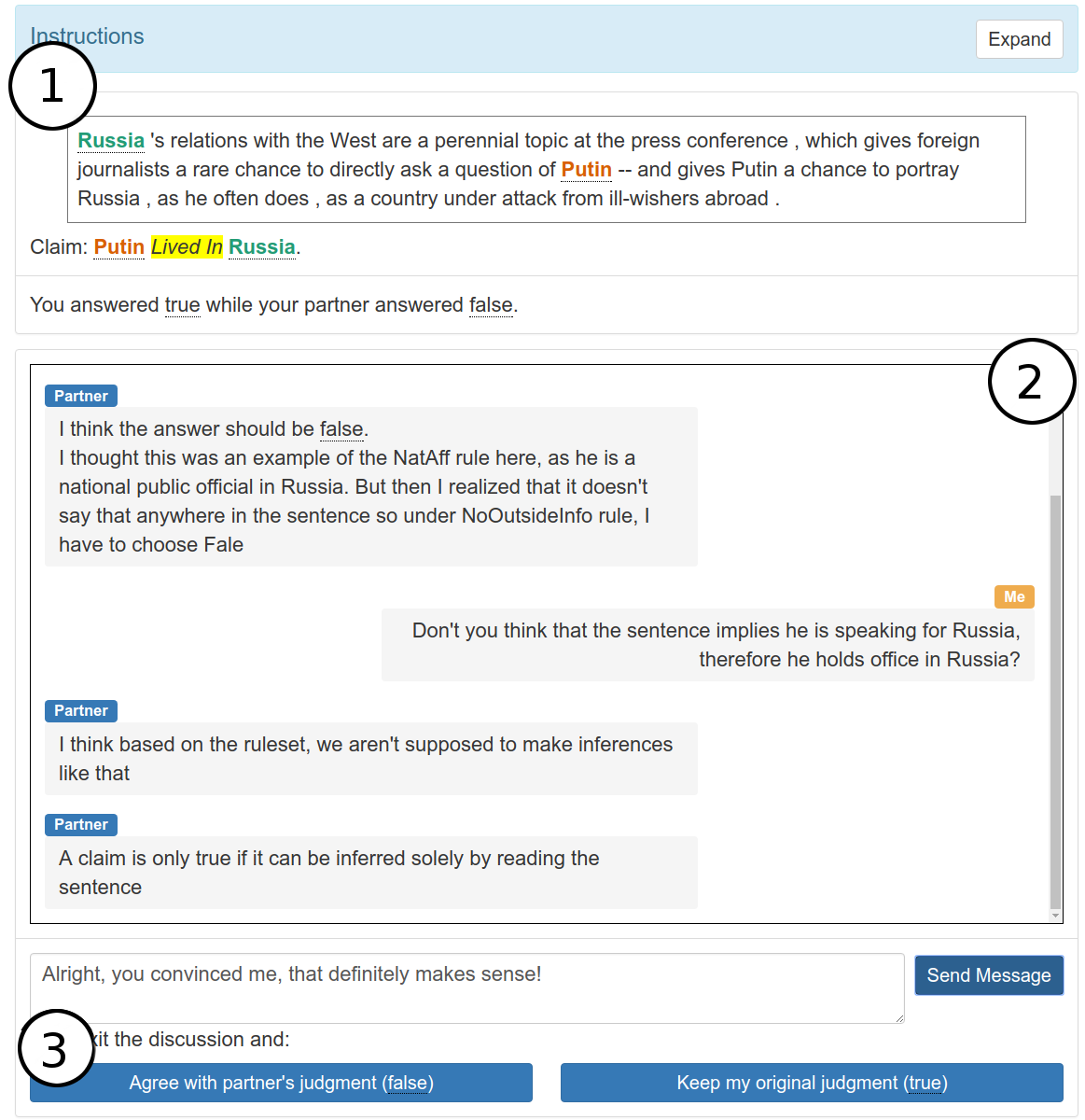}
 \caption{Discussion interface for use in \sys, inspired by instant-messaging clients, showing fragment of an actual discussion in the Relation Extraction domain. (1) Presents the question (sentence + claim) and both sides' beliefs. (2) Initial discussion is seeded with the workers' justifications. (3) Options added to facilitate termination of a discussion once it has reached the end of its usefulness.}~\label{fig:discussion_screenshot}
  \vspace{-0.3in}
\end{figure}

Yet recently, {\em argumentation} has been shown to be an effective way to improve the accuracy of both individual and aggregate judgments. For example, Drapeau \etal's MicroTalk~\cite{Drapeau2016MicroTalkUA} used a pipelined approach of: 1) asking crowdworkers to {\em assess} a question's answer, 2) prompting them to {\em justify} their reasoning, 3) showing them counterarguments written by other workers, and 4) allowing them to {\em reconsider} their original answers to improve individual judgments. In principle, this simplified form of argumentation allows a single dissident worker, through force of reason, to steer others to the right answer. Furthermore, the authors showed that argumentation was compatible with EM; combining the two methods resulted in substantial gains in accuracy.

However, while asynchronous argumentation systems like MicroTalk attempt to resolve disagreement, the steering power of a one-round debate is limited. Workers are only shown a pre-collected justification for an opposing answer; they aren't challenged by a specific and personalized argument against the flaws in their original reasoning. There is also no back-and-forth interaction that could illuminate subtle aspects of a problem or resolve a worker's misconceptions --- something which may only become apparent after several turns of discussion. Furthermore, since justifications are pre-collected, workers need to write a generic counter argument; while this works for binary answer tasks,  it is completely impractical for tasks with many answers; such a counter-argument would typically be prohibitively long, refuting $n-1$ alternatives.  

This paper presents \sys, a new workflow that engages workers in {\em multi-turn and contextual} argumentation to improve crowd accuracy on difficult tasks. \sys\ selects workers with opposing answers to questions and pairs them into a discussion session using a chat-style interface, in which they can respond to each other's reasoning and debate the best answer. During these exchanges, workers are able to write context-dependent counter-arguments addressing their partner's specific  claims, cite rules from the training materials to support their answers, point out oversights of other workers, and resolve misconceptions about the rules and task which can impact their future performance on the task. As a result of these effects, workers are more likely to converge to correct answers, improving individual accuracy. Our experiments on two difficult text based task domains, relation extraction and a word association task, show that contextual multi-turn discussion yields vastly improved worker accuracy compared to traditional argumentation workflows. 

In summary, we make the following contributions:

\begin{itemize}
\item We propose \sys, a novel workflow that induces multi-turn and contextual argumentation, facilitating focused discussions about the answers to objective questions. 

\item We introduce a new type of worker training to ensure that workers understand the process of argumentation (in addition to the task itself) and produce high quality arguments.

\item We develop \implement, a synchronous implementation of our workflow using real-time crowdsourcing, and apply it to conduct the following experiments: 

    \begin{itemize}
    \item In the Relation Extraction domain introduced by  {\sc MicroTalk}~\cite{Drapeau2016MicroTalkUA},
    we show that contextual, multi-turn argumentation results in significantly higher improvement for individual worker accuracy over existing one-shot argumentation: a 16.7 percentage point improvement  {\em v.s} 6.8.

    \item Using a version of the Codenames domain~\cite{Zhou2018InSO}, that has many answer choices (making {\sc MicroTalk}'s non-contextual argumentation unworkable), we show that \sys\ is quite effective, improving individual worker accuracy from 66.7\% to a near-perfect 98.8\%.

    \item We qualitatively analyze the discussion transcripts produced from our experiments with \implement, identifying several characteristics present in contextual, multi-turn argumentation.
    \end{itemize}

\end{itemize}

\begin{figure*}
  \centering
  \includegraphics[width=1.9\columnwidth]{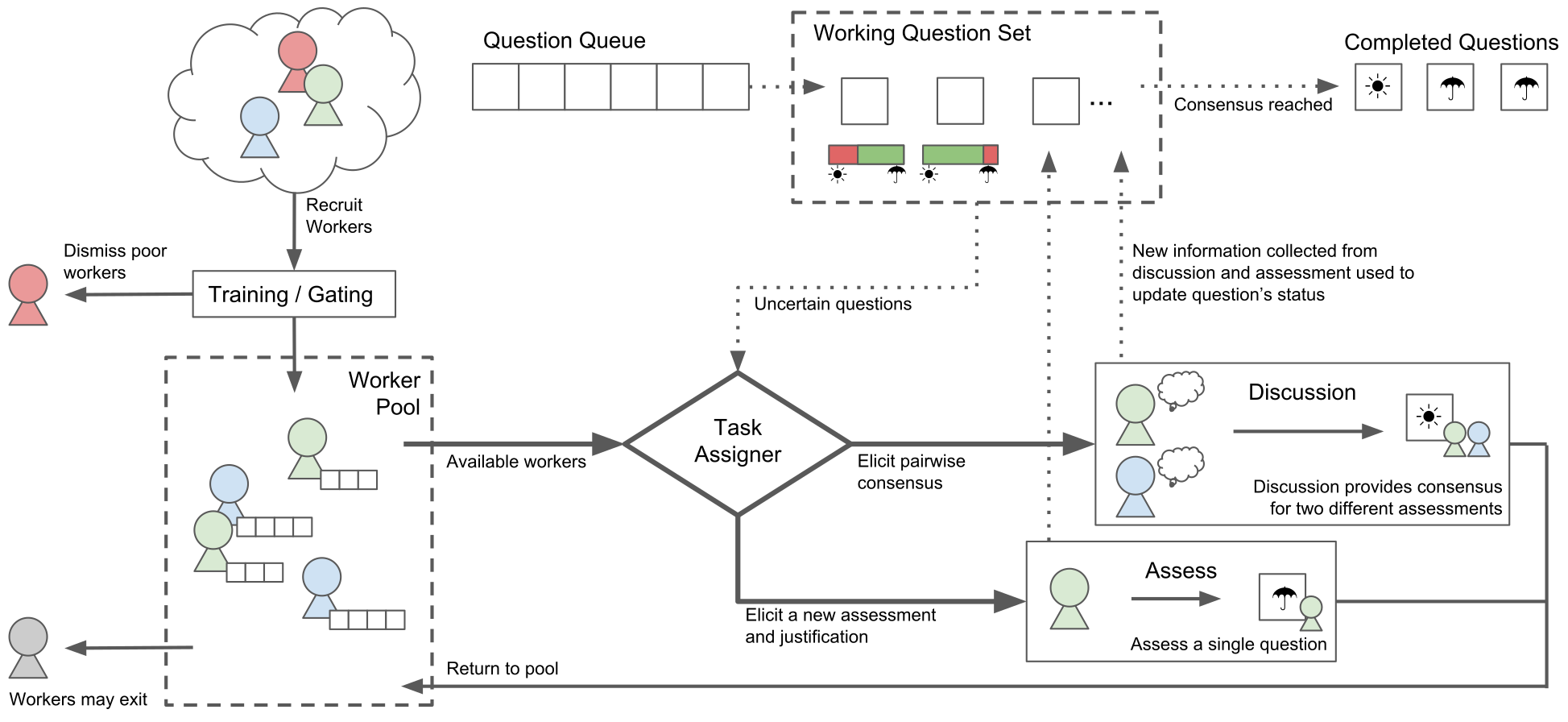}
  \caption{\sys\ System Diagram. Solid arrows indicate paths for workers through the system. Dotted arrows indicate how questions are allocated through the system. }~\label{fig:system_diagram}
\end{figure*}

\section{Previous Work}

Quality control has been a central concern in crowdsourcing, and space precludes a complete discussion of post-hoc methods such as majority vote~\cite{snow-emnlp08}, gated instruction~\cite{liu-naacl2016}, and programmatic generation of gold-standard questions~\cite{oleson-hcomp2011}. Similarly, many different approaches have been suggested to improve accuracy by assigning certain question to specific workers~\cite{karger2011,dai-aij13,kamar-aamas12}.

Expectation maximization~\cite{dawid-as79,whitehill-nips09,welinder-nips10} is especially popular, but all methods embody greedy optimization and hence are prone to local maxima. If the data set includes challenging problems, where a significant majority of workers gets the answer wrong, EM will likely converge to the incorrect answer. 

Other researchers have investigated methods to handle cases where the majority may be wrong, \eg, \ tournament voting~\cite{sun-nips11} and Bayesian truth serum~\cite{prelec-objective-bts}. Unfortunately, these methods are unlikely to work for difficult questions where individual answers require detailed analysis.  

\subsection{Rationales \&\ Feedback Can Improve Accuracy} 

Some researchers have demonstrated that requiring annotators to submit ``rationales'' for their answers by highlighting portions of text~\cite{zaidan-naacl07,mcdonnell-hcomp16} or an image~\cite{donahue-iccv11} can improve machine learning classifiers. In contrast, we focus not just on providing justifications, but also back-and-forth dialog between workers. 

Dow \etal~\shortcite{dow-cscw12} conduct experiments demonstrating that timely, task-specific feedback helps crowd workers learn, persevere, and produce better results. Wiebe \etal~\shortcite{wiebe-acl99} go a step further (with small-group, in-person studies), showing the benefits of getting annotators to reconsider their positions and discuss them with other workers. ConsiderIt~\cite{Kriplean2011ConsiderItIS} takes this kind of principled debate online and into the political spectrum, using pro/con points and encouraging participants to restate alternative positions to help voters make informed choices. Recently, Kobayashi \etal~\cite{Kobayashi2018AnES} have shown that self-correction in the form of reviewing other workers' answers is effective in getting workers to reconsider and correct their own.

Many of these papers inspired the {\sc MicroTalk} system, which is the primary inspiration for this work~\cite{Drapeau2016MicroTalkUA}. {\sc MicroTalk} combines three microtasks --- assess, justify \&\ reconsider --- asking crowd workers to assess a question, prompting them to justify their reasoning, confronting them with another worker's counterargument, and then encouraging them to reconsider their original decision. Our work builds on {\sc MicroTalk} in several ways: 1) we support contextual communication that allows participants to respond to specific points made by their partner, and 2) we support multi-turn dialog that allows workers to ``dive deep'' if necessary to resolve the disagreement. 

Liu \etal~\shortcite{Zhu:2014:RVD:2531602.2531718} also noted that workers who review others' work perform better on subsequent tasks, which inspired us to also examine effects of multi-turn argumentation on workers' future tasks. 

\subsection{Real-Time Crowdsourcing}

In order to facilitate synchronous and real-time dialog, our implementation, \implement, has to connect workers in real-time. There are significant logistical challenges with recruiting a real-time crowd~\cite{huang-hcomp16}, especially in the context of our experiments, where we wished to have several dozen workers working simultaneously. Fortunately, there have been enough real-time, crowd deployments (\eg,~\cite{bernstein2010soylent,bigham-uist10}) that many useful lessons have been distilled~\cite{huang-hcomp17}.

\subsection{Instructions for the Task \&\ the Meta-Task}

It's generally agreed that good instructions are essential for high inter-annotator agreement~\cite{liu-naacl2016}. 
We were frequently reminded of this as we iterated on our task design. Perhaps we should not have been surprised to discover that good instructions were also necessary for our `meta-task' --- arguing about the task. Argument forms and norms that contribute to positive discussion have been long-studied in the education community, termed `accountable talk'~\cite{michaels2008deliberative}. Our training for each experiment domain instructs workers on some of these guidelines, \eg\ asking workers to rate justifications to ensure that participants understand accepted standards of reasoning, arguments that emphasize logical connections, and the ability to draw a reasonable conclusion. 

\section{\unscsys\ Design}
We designed the \sys\ workflow to address issues in existing crowdsourced argumentation systems by using contextual, multi-turn discussions to address drawbacks in one-shot reconsider systems. 

In this section, we present the \sys\ workflow as well as design considerations in a synchronous implementation of the workflow used for our experiments. We first explain the rationale for contextual, multi-turn discussions and give an overview of our \sys\ workflow. We then talk about the decision to implement our workflow in a synchronous system---\implement. Finally, we discuss the design choices we made to (1) create an interface for effective real-time discussion, as well as (2) improve instructions and training for the domains we examined.

\subsection{Contextual and Multi-Turn Discussion}
In natural forms of debate, participants who disagree take turns presenting arguments which can refute or supplement prior arguments. Our \sys\ workflow is designed around the concept of emulating this process in a crowd work setting by using paired discussions facilitated by a dynamic matching system. Participants are matched with partners based on their current beliefs and are encouraged to present their arguments over multiple turns. 

While real-life debates may include multiple participants each responsible for addressing arguments on different aspects of a problem, in the crowd setting we can utilize the diversity of workers to cover a broad set of views and reasoning; thus, to simplify the process, we focus on a two-participant discussion model.

\subsection{Workflow Overview}

Since argumentation happens on an ad-hoc basis, it's much more flexible to have our workflow focus on managing transitions between different states a worker may be in instead of defining a single pipeline. Due to this, our design of the \sys\ workflow follows an event-based definition model where the task assigner allocates tasks as workers' state changes. Figure~\ref{fig:system_diagram} summarizes how our workflow allocates worker resources and questions in a dynamic way. 

Initially, workers are recruited from a crowd work platform (such as Amazon Mechanical Turk) and are immediately assigned to a \textbf{training} task. Workers who pass training and the associated gating tests~\cite{liu-naacl16} enter the \textit{worker pool} and wait to be assigned work. Then, instead of a fixed workflow, our event-based task allocator decides which type of task and question to assign to a worker subject to a set of constraints. As workers complete their tasks and update the beliefs of questions in the working set, new candidate tasks are dynamically selected and allocated. 

In \sys, there are two main types of tasks that the assigner may assign to an idle worker: \textbf{assess} and \textbf{discussion}.

\begin{itemize}
  \setlength{\itemsep}{1pt}
  \item The \textbf{assess} task acquires one worker ($w$) from the worker pool who is then presented with one question ($q$) --- in our case a single question in the  domain --- that asks for an answer to a multiple choice question ($\text{Belief}_w(q)$) and optionally a free-form justification for their position ($\text{Justification}_w(q)$). This task is a combination of the assess and justify microtasks in MicroTalk~\cite{Drapeau2016MicroTalkUA} as a single task.
  \item The \textbf{discussion} task acquires two workers ($w_1, w_2$) from the worker pool who are both shown a discussion interface for a question ($q$). At the end of a discussion, $\text{Belief}_w(q), \text{Justification}_w(q)$ may be updated for $w \in \{w_1, w_2\}$. We will cover details on the design of the discussion task in later sections.
\end{itemize}

The task allocator manages which type of task can be allocated when a worker changes their state and, depending on domain, can be adjusted to prioritize specific kinds of tasks, particular questions or qualities such as minimizing worker wait time and increasing concurrent work. 

In general, the task assigned can be adapted to the goals of the requester. However, there are a few general constraints that the task assigner must follow:

\begin{itemize}
  \setlength{\itemsep}{1pt}
  \item \textbf{Incompatible beliefs}: A discussion may only be assigned to workers ($w_1, w_2$) if they have incompatible beliefs ($\text{Belief}_{w_1}(q) \neq \text{Belief}_{w_2}(q)$). Implicitly, this also requires existence of the both beliefs, implying they must have been collected.
  \item \textbf{No repeated discussions}: Two workers ($w_1, w_2$) may only discuss question $q$ if they have never discussed question $q$ with each other before ($\lnot \exists A^{\text{Discuss}}_{q}(w_1, w_2)$).
\end{itemize}

These constraints guarantee that the workflow will eventually terminate when there are no more workers who disagree and have never paired with each other. It's important to note that requesters can set up the task allocator to terminate the workflow sooner if, for example, thresholds for agreement on questions are reached.

\subsection{\implement: A Real-Time Implementation}

While the \sys\ workflow does not constrain the type of interaction during a discussion task, we decided to test out the effectiveness of our workflow using synchronous discussions where both workers are simultaneously online and engaged in a chat-like discussion environment. 

In this implementation of \sys, \implement, once workers are matched into a discussion, they will not be assigned other tasks for the duration of that discussion and are expected to give each other their undivided attention. A synchronous and real-time discussion task allows us to maintain a continuous dialogue spanning many turns while preserving discussion context in a simple and natural way. However, we came to learn that systems relying on synchronous, real-time worker interactions have some disadvantages: The synchronous nature of discussions means that some workers will have to wait for a partner to become available and workers need to be online and active within the same time window, both of which  imply a higher cost to the requester. 

Additionally, there are many practical challenges to implementing and setting up synchronous real-time experiments with crowd workers, including implementing real-time client-server communication and working with APIs for worker recruitment and payment. We elected to build \implement\ on top of the TurkServer~\cite{Mao2012TurkServerES} toolkit. Various tools built into TurkServer simplify the interfacing with Amazon Mechanical Turk for worker recruitment and task management and allow us to automatically track worker state as well as building a ``worker pool'' through the \textit{lobby}. These allowed us to quickly design and prototype \implement, which builds upon TurkServer's lobby-assigner-experiment model. Our \textit{training}, \textit{assess}, and \textit{discussion} tasks types in \implement\ map experiment instances in TurkServer.

\subsection{Discussion Interface}

The discussion task is the most important and defining task of the \sys\ workflow. We considered multiple different options for the discussion interface focusing on ways to organize discussion structure and facilitate discoverability. 

Early proposals included designs that were inspired by the posts-and-replies interfaces in social network timelines and the split-view pros-and-cons interfaces used in ConsiderIt, a political, argumentation system~\cite{Kriplean2011ConsiderItIS}. Our pilot studies showed that these methods were cumbersome and non-intuitive, so we decided on a free-form instant messaging (chat) metaphor for the discussion task; shown in Figure~\ref{fig:discussion_screenshot}. 

When a pair of workers enter a discussion, they are placed into a familiar instant messaging setting, where they can freely send and receive messages. An additional ``exit'' section below the chat interface allows either participant to terminate the discussion if they feel that it is no longer useful. Workers can utilize this exit mechanism to indicate that a consensus was reached or that no agreement is possible between the pair. The discussion interface can be easily adapted to specific needs of each experiment domain: In the Relation Extraction domain, the justifications collected from earlier assess or discussion tasks are used to seed the system, which we found to be beneficial in starting a conversation. In the Codenames domain, a drop-down menu below the text input field accommodates switching to alternate answers during the discussion addressing the non-binary nature of the questions. 

We found that workers required minimal training to understand the discussion interface and were quickly able to effectively participate in argumentation with their partners in both domains. 

\subsection{Optimizing Task Instructions}
\begin{figure}
  \centering
  \includegraphics[width=1\columnwidth]{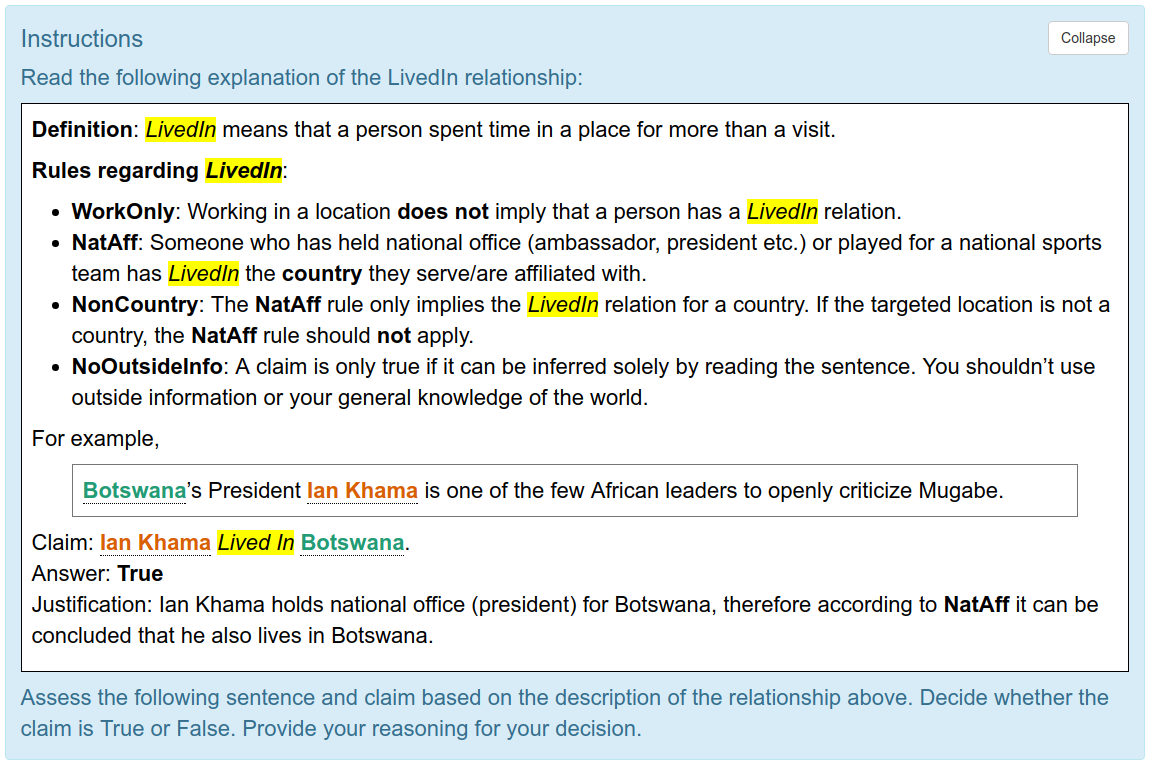}
  \caption{Screenshot of our \textit{LivedIn} task (Relation Extraction domain) instructions containing 5 citable rules including the definition. Shorthands (in bold) allow for efficient citation of rules during discussion and within justifications (as shown in the example's justification).}~\label{fig:instructions_screenshot}
  \label{fig:instructions}
  \vspace*{-0.15in}
\end{figure}

We observed in early pilot experiments that arguments which refer explicitly to parts of task guidelines were more effective at convincing a partner. However, our original task guidelines and training did nothing to encourage this practice. Workers came up with different ways to refer to parts of the instructions or training examples, but this was inconsistent and frequently caused confusion. References to the guidelines were hard to identify making it harder for workers to determine correct invocations of rules in the Relation Extraction domain pilots. Since arguing in synchronous discussion sessions is time-sensitive, creating rules and shorthands that are easy to cite is important for discussion efficiency.

We adjusted the task guidelines for the Relation Extraction task from those in {\sc MicroTalk}, re-organizing them into five concrete and easy-to-cite rules as shown in Figure~\ref{fig:instructions_screenshot}. Each rule was given a shorthand so that workers can unambiguously refer to a specific rule and aid in identification of proper or improper rule usage during the discussions. We observed that citing behavior became more consistent within discussions with workers frequently utilizing our shorthands.

In the Codenames domain, which has simple instructions but a lot of example cases, we designed the instructions to both show the general guidelines and also provide a way for workers to review examples from training if they decide to reference them. 

\subsection{Selecting and Training Effective Workers}

In initial pilots with \implement, we provided minimal training to workers. New workers were presented with task descriptions, instructions and one sample problem which could be attempted multiple times. After several pilot experiments, we noticed that workers were performing inconsistently. Many workers' discussions were ineffective, with one or both showing lack of basic understanding of the instructions. Inspection of timing data also showed that many workers were spending significantly less time in training and assess tasks without putting the necessary thought into learning the rules.

One method for improving worker quality, suggested by Drapeau \etal~\cite{Drapeau2016MicroTalkUA}, is to select for ``discerning workers'' by finding workers who write justifications that are more complex than those of the median worker \eg, using Flesch-Kincaid~\cite{flesch-kincaid} readability grade levels on gold standard questions. Following Drapeau \etal, we tried filtering for ``discerning workers'' using the Flesh-Kincaid score but our pilot experiments on the Relation Extraction domain showed limited effect and significant reductions in worker pool size. Filtering of workers based only on gold standard question performance was also ineffective as it did not train workers to understand the rules required for our complex tasks.

Since our tasks require worker training, we opted to implement a gating process~\cite{liu-naacl2016}, that can both train and select workers at the same time. Workers are presented with questions laid out in a quiz-like format. Each training question is provided along an introduction of related concepts from the task instructions. The questions are interleaved with the instructions so that new questions are presented as new concepts are introduced to reinforce the understanding. Feedback is given when a workers selects an answer. At the end, workers' performance on a set of quiz questions is recorded. If a worker's accuracy on the quiz falls below a certain threshold, the worker will be asked to retry the training section (a limited number of times) with the order of the quiz questions randomized. Workers are dismissed if they exceed the retry limit and still do not meet the passing threshold. 

After the improved training and gating, we observed no significant correlation between workers' initial accuracy and the ``discerning worker'' measure in the Relation Extraction domain. 

\subsection{Selecting and Training Effective Argue-ers}

In addition to gating for the task instructions, we also designed a novel {\bf justification training} task incorporated as a part of the training process with the goal being to train the workers on recognizing good arguments and justifications. This justification training task is presented in the form of an assess task followed by providing feedback to workers explaining good arguments and justifications based on their assessment. 

In the Relation Extraction domain, workers are asked to identify the {\em best} justification in a list after providing an assessment. Incorrect options aim to address potential mistakes a worker may make in writing a good justification, such as: failure to cite rules, incomplete or incorrect references to the rules, or making extended and inappropriate inferences. Workers are given feedback explaining why their selected justification was incorrect or correct with both the good and bad parts of the justification explained. 

Since questions in the Codenames domain have ten or more possible answers, it's not practical to collect justifications (which would have to rule out $n-1$ alternatives) at assess time. Therefore, justification training is adjusted to instead show a context-dependent counter-argument when a worker selects an incorrect answer refuting the incorrect choice and supporting the correct one. By training workers to recognize and analyze arguments, the goal of justification training is to promote more critical discussion. 

\subsection{Worker Retention and Real-Time Quality Control}

Due to the synchronous nature of discussions in \implement, workers may become idle for short periods of time when they are waiting to be matched to a partner. In these circumstances, workers are kept in the {\em worker pool} in the form of a {\em lobby}. Our lobby design was mainly inspired by both the default lobby provided in TurkServer~\cite{Mao2012TurkServerES} and from a worker-progress feedback design developed by Huang \etal~\cite{huang-hcomp17} for low-latency crowdsourcing. While in the lobby, workers are presented with information on their peers' current status, such as how many workers are currently online and which workers may become available soon. Workers also see statistics on their work, which is tied to bonus payments, and are encouraged to wait. In \implement, the task assigner is configured to immediately assign work as it becomes available, but we also allow a worker to voluntarily exit after they have waited for a period without matches or completed a sufficiently large amount of tasks. 

In addition, while our gating process is designed to select workers serious about the task, we do incorporate several techniques to assure that workers stay active when a task gets assigned to them. Individual tasks, such as assess tasks, impose anti-cheating mechanisms to discourage spammers from quickly progressing. These mechanisms include character and word count minimums and disabling of copy-paste for free-form entries. Workers are also encouraged to peer-regulate during discussion --- participants can indicate a partner's inactivity upon ending a discussion with no agreement. Paired with corresponding payout incentives, these methods ensure that most workers stay active throughout the duration of an experiment.

\newcommand{\numposttest}{4}

\section{Experiments}

We deployed our experiments on our synchronous implementation, \implement, to address the following questions: 1) Does multi-turn discussion improve individual accuracy more compared to existing one-shot reconsider based workflows? 2) Is multi-turn discussion effective in cases where acquiring justifications to implement one-shot argumentation (reconsider) is impractical?  3) Do discussions exhibit multi-turn and contextual properties?

We selected two domains to evaluate the research questions above: a traditional NLP binary answer task, Relation Extraction, for comparing against one-shot argumentation and a multi choice answer task, inspired by the word relation game Codenames, to evaluate \sys\ in a non-binary choice domain. 

In the following sections, we first introduce the experiment setup and configuration, then we introduce each domain and present our results for experiments on that domain. At the end, we present a qualitative analysis of discussion characteristics and explore whether discussions can improve future accuracy.

\subsection{Experiment Setup}

\sys's design enables interleaved assignment of different task types (assessments or discussions) for individual workers. This can be beneficial in reducing worker waiting overhead by assigning individual tasks when paired tasks are not available. However, in order to evaluate the effects of contextual, multi-turn argumentation under a controlled setting, we need to isolate the process of assessment and argumentation. For our experiments, we implemented a ``blocking'' task assigner that avoids interleaved concurrent tasks and is designed to assign the same type of task to a worker until they have answered all questions of that type. 

The \textit{blocking assigner} includes a few extra constraints in addition to those required by the workflow:

\begin{itemize}
  \setlength{\itemsep}{1pt}
  \item \textbf{Gold Standard Assessments}: The task assigner assigns \textbf{assess} tasks for gold standard questions to evaluate quality of workers who passed the training and gating quiz phase. Workers are assigned these questions before any other questions. No discussions are ever initiated for these questions; they let us control for worker quality and filter workers that do not pass the gating threshold. 
  \item \textbf{Greedy Matching}: The task assigner tries to assign a discussion as soon as such a task is available. In the case of multiple candidates, the task assigner picks one randomly.
\end{itemize}

Additionally, the \textit{blocking assigner} doesn't allocate any discussions until a worker has finished \textit{Assess}-ing all questions. This allows us to collect the initial answers of a worker before they participate in any argumentation. 

We adjusted \implement\ to include these experimental constraints and collected multi-turn, contextual arguments. The system used in experiments consists of three distinct stages: \textit{Training}, \textit{Assess} and \textit{Discussion}. Workers progress through each stage sequentially. Within the same stage, workers will be allocated tasks on demand by the \textit{blocking assigner}. This system represents the \textbf{discussion} condition.

In addition to \implement, we also implemented the adaptive workflow described in MicroTalk~\cite{Drapeau2016MicroTalkUA} using our platform to compare with one-shot argumentation. We reproduced the \textbf{reconsider} task interface from MicroTalk which replaces our \textbf{discussion} task and created a specialized assigner that allocates tasks following the adaptive algorithm described in MicroTalk. In this system, workers will be adaptively asked to justify or do \textbf{reconsider} tasks depending on their initial answer. When a worker is the only worker with a particular answer for a question, they will be asked to provide a justification for their answer. \textbf{Reconsider} tasks are only assigned to a worker if there is a previously justified answer that their current answer disagrees with. That justification will be the one shown to the worker. This system represents the \textbf{reconsider} condition.

\subsection{Recruiting and Incentives}
We ran experiments on Amazon Mechanical Turk, using workers who had completed at least 100 tasks with a 95\% acceptance rate for both of our experiment domains. 

Within each domain, we calibrated our subtask payments by observing the average worker time for that subtask from a pilot run and allocating an approximately \$7 hourly wage. Our training bonus of \$1.00 for successfully completing training and the gating quiz is also calibrated using the average time it takes workers to complete the training session. 

For the Relation Extraction domain, workers are paid \$0.10 as base payment and \$1.00 for training. Workers are paid a per-question bonus of \$0.05 for an assessment, \$0.05 for a justification, and --- depending on their condition --- a bonus of \$0.50 for a \textbf{discussion} or \$0.05 for a \textbf{reconsider} task. Since we always collect a justification for each question in the \textbf{discussion} condition, workers in that condition are always given the full \$0.10 per-question bonus. Per-question incentives are chosen to match those used in MicroTalk~\cite{Drapeau2016MicroTalkUA}.

For the Codenames domain, workers are paid \$0.20 as base and \$1.00 for training. Workers are paid a per-question bonus of \$0.20 for each correct answer and a per-discussion bonus of \$0.50 for participating in a discussion with an extra \$0.25 for holding the correct answer at the end of that.

While it is possible to design a more complex incentive structure, our main goal for this set of incentives is to discourage cheating behavior and align with that of MicroTalk. We think these incentives are consistent with those used in other, recent crowdsourcing research~\cite{liu-naacl2016}. 

\subsection{Relation Extraction Domain: Binary Answer}

\begin{figure}
  \centering
  \includegraphics[width=1.0\columnwidth]{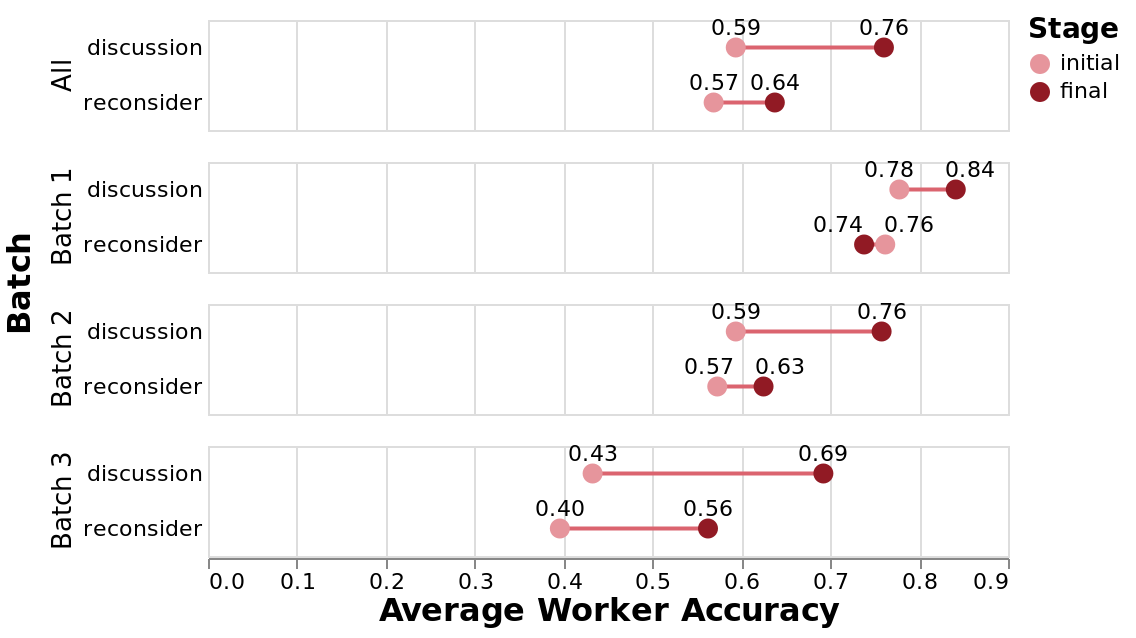}
  \caption{Comparison for improvement in average worker accuracy (Relation Extraction domain) for each batch (subset) of questions as well as on the entire set of questions.}~\label{fig:plot-relation-accuracy}
  \vspace*{-0.15in}
\end{figure}

In the interest of comparing to previous work, we evaluated our method on a tradition NLP annotation task of {\em information extraction} (IE) --- identifying  structured, semantic information (relational tuples, such as would be found in a SQL database) from natural language text~\cite{grishman-ie-1997}. The task is of considerable interest in the NLP community, since most IE approaches use machine learning and many exploit crowdsourced training data~\cite{zhang-acl2012,pershina-acl2014,angeli-emnlp14a,liu-naacl2016}.

Specifically, we consider the problem of annotating a sentence to indicate whether it encodes the TAC KBP  {\it LivedIn} relation ---
does a sentence support the conclusion that a person lived in a location? While such a judgment may seem  simple, the official LDC annotation guidelines are  deceptively complex~\cite{surdeanu-kbp13}. For example, one can conclude that a national official lives in her country, but not that a city official lives in her city. Figure~\ref{fig:instructions} defines the task, showing the instructions given to our workers.

We created a set of 23 challenging TAC KBP questions drawing from the 20 used in MicroTalk~\cite{Drapeau2016MicroTalkUA} and adding 3 additional questions from Liu \etal~\cite{liu-naacl2016}. This set was then divided into 3 batches of size 7, 8, and 8 for our discussion experiments. For gold standard questions, we selected 3 simple questions from the TAC KBP set, each of which can be resolved with an invocation of one rule. Upon recruitment, each worker is also presented with a 6 question gating quiz and are allowed 2 attempts to pass the gating threshold. Gating questions were written to be simple and unambiguous, testing whether the worker was diligent and had absorbed the guidelines.

\subsection{Multi-turn {\em vs.} One-shot Workflows}

Our first experiment compares worker accuracy for the multi-turn, contextual discussion workflow design against that of an one-shot (non-contextual) reconsider workflow on the binary answer Relation Extraction domain (\ie, \sys\ {\em vs.} {\sc MicroTalk}). We deployed both conditions with the systems described in the experiment setup. Since workers need to complete all assessments before starting discussions which would cause increased waiting time on a large set of questions, we deployed the {\em discussion} condition experiments in 3 batches (N=9, 16, 13) corresponding to the 3 batches the experiment questions were divided into. In the {\em reconsider} condition (N = 12), workers were put through our implementation of the adaptive workflow from {\sc MicroTalk} on all questions. 

In both conditions, the gating threshold was set at $100\%$ --- workers needed to answer all gold standard questions correctly to be included. From the plot shown in Figure~\ref{fig:plot-relation-accuracy} we can see that the {\em discussion} condition (\sys) improves average worker accuracy by $16.7$ percentage points over the initial accuracy compared to $6.8$ for the {\em reconsider} condition (statistically significant at $p=0.0143$). 

We performed a t-test on the initial accuracy of workers across both conditions for each batch and found no statistically significant difference ($p = 0.77, 0.78, 0.67$) indicating that workers of similar quality were recruited for each of our batches.
On average, workers participated in $7.7$ discussions ($\sigma = 4.75$) and were presented with $16.8$ reconsider prompts ($\sigma = 3.83$) in the one-shot workflow. 

We do note that discussions are more costly, largely due to paying workers for time spent waiting for their partner to respond.  Each \implement\ discussion took an average of $225.3$ seconds ($\sigma = 234.8$) of worker time compared to a one-shot reconsider task averaging $13.6$ seconds ($\sigma = 15.0$). We believe that an asynchronous implementation of \sys\ could reduce overhead and dramatically lower costs.

\subsection{Codenames Domain: Multiple Choice with Many Answers}

\begin{table}
  \centering
  \begin{tabular}{|r|l|}
    \hline
    Candidates & {business, card, knot} \\
    \hline
    Positive Clues &  {suit, tie} \\
    \hline
    Negative Clues &  {corporation, speed} \\
    \hline
    Explanation & \makecell[tl]{
     Workers must find the single best \\
     candidate word that is related in meaning\\
     to some positive clue word, but none of \\
     the negative clues. In this example, \\
     all three candidates are related to \\
     some positive clue: a suit \\ 
     for business, a suit of cards, and to tie a \\
     knot. However, business relates to \\
     corporation and knot is a unit of speed. \\
     Card is the best answer: it's related to \\
     a positive clue, while being largely \\
     unrelated to any negative clues.} \\
    \hline
     Best Answer &  {card} \\
    \hline
  \end{tabular}
  \caption{Example of a simple question used for training from the Codenames domain. Real questions have around 7-10 candidate words.}
  \label{tab:codenames-example}
  \vspace*{-0.15in}
\end{table}

Previous work using one-shot argumentation focused mainly on evaluating argumentation in domains that only acquired binary answers such as Relation Extraction. However, we observed that this is not sufficient to represent a wide variety of real world tasks. In a binary answer setting, workers are able to write justifications arguing against the single opposing belief. As the number of answer options grows, it becomes increasingly inefficient and even infeasible to collect justifications as a part of each worker's assessment. 

Effective justifications for multiple choice answers need to address not only the selected answer, but also argue against remaining options, making them long and difficult to understand. Multi-turn discussion can address these scaling issues through back-and-forth dialog through which workers can argue against their partner's specific choice.

We created a synthetic task inspired by the popular word association {\em Codenames} board game as a multiple choice answer domain. The Codenames game has been adopted as the main task in other cooperative work designs such as DreamTeam~\cite{Zhou2018InSO}, which utilized a cooperative version of the game. The objective in the game is for each team to identify the tiles assigned to them from a shared list of word tiles. Clue words are given by one team member (the ``spymaster'') who can see the assignment of word tiles (which ones belong to which team) while other teammates have to find the correct word tiles for their team while avoiding the tiles assigned to the other team. 

Our Codenames task domain draws inspiration from the competitive aspect of the game. We observe that late into the game, good word guesses are often informed by both the teammate clues (which should be matched) and opponent clues (which should be avoided). With this observation, we created tasks which consist of a list of candidate words, several positive and several negative clue words. Workers, in the role of a team member, are instructed to find the single best candidate word that is related in meaning to some positive clue word but none of the negative clues. An example of this task can be seen in Table~\ref{tab:codenames-example}. Each question contains around 2 positive clues, 2-3 negative clues and 7-10 candidate words. We created 3 gating questions, 7 experiment questions and 1 individual assessment question for this task. We used a gating threshold of $66.7\%$. While Codenames is not a typical task for crowd work, as also noted in DreamTeam, we think its aspect of multiple choice answers is representative of a whole class of similar tasks that lack effective one-shot argumentation strategies. 

The loose definition of words being ``related'' in Codenames domain reduces the amount of worker training required for participation since it utilizes common knowledge of language. However, this may lead to ambiguity in reference answers which would be undesirable. We elected to manually create a set of questions which were validated to have only 1 objectively best answer. The distractors for each question and our reference argument were evaluated with a group of expert pilot testers. We confirmed that all participants agreed with our reference counter-arguments against the distractors and also with our reference answer. In the pilot test, we also noted that this task can be very challenging even for experts as multiple word senses are involved in distractors. 

\subsection{Evaluating on Multiple Choice Tasks with Many Answers}

\begin{figure}
  \centering
  \includegraphics[width=1\columnwidth]{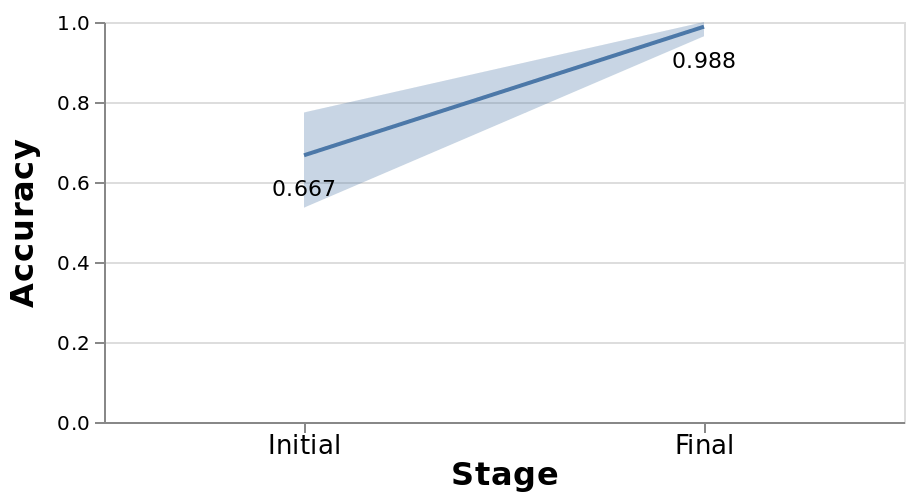}
  \caption{Initial and final accuracy of multi-turn argumentation on the Codenames domain with 95\% confidence intervals.}
  \label{fig:plot-codenames-accuracy}
  \vspace{-0.15in}
\end{figure}

Our second experiment examines the performance of \implement\ on multiple choice answer tasks from the Codenames domain, a domain that would be very inefficient for one-shot argumentation (justifications would need to address up to 9 alternatives). While initial worker accuracy is only $66.7\%$, 
\implement\ achieves a final average worker accuracy of $98.8\%$, a significant improvement 
($N=12$). 

We performed the ANOVA omnibus test with a mixed effects model using worker initial accuracy as a random effect and found that the experiment stage (initial {\em v.s.} final) is statistically significant at ($F(1, 57.586) = 85.608, p = 5.445 x 10^{-13} < 0.001$). The average duration of each discussion was $123.56$ seconds ($\sigma = 64.79$) and workers had an average of $6.3$ discussions ($\sigma = 3.89$) in the Codenames domain. 

\subsection{Discussion Characteristics}

We can see from the previous experiments that multi-turn, contextual argumentation is effective at improving worker accuracy across a variety of tasks, but are the discussions actually taking advantage of multi-turn arguments and the context being available? To answer this question, we collected and analyzed the transcripts recorded for each domain: Relation Extraction and Codenames.

We computed statistics on multi-turn engagement by analyzing the number of worker initiated messages -- each of which is considered a turn. We found that in the Relation Extraction domain, discussions averaged $7.5$ turns ($\sigma = 6.1$, median of $5$) while in Codenames discussions averaged $8.3$ turns ($\sigma = 4.23$, median of $7$). We also found that in  Codenames, the number of turns correlates to convergence on the correct answer ($F(1, 31) = 7.2509, p < 0.05$) while we found no significant relation between turns and convergence ($p > 0.1$) in the Relation Extraction domain. We note that in Relation Extraction, discussions are seeded with workers' justifications from the assess task (equivalent to 2 non-contextual turns, not counted in the average) whereas discussions in the Codenames domain use actual contextual turns to communicate this information. Codenames discussions also utilized extra turns to argue about alternative choices.

Additionally, we noticed several patterns in the discussion text that appeared in both domains. We further examined these patterns by coding the the discussion transcripts (147 from Relation Extraction and 38 from Codenames). We surveyed the discussions looking only at patterns specific to argumentation and came up with 8 patterns related to argumentation techniques and 6 reasons workers changed their answer. 

We then narrowed down the argumentation patterns by removing any that were highly correlated or any that had just 1-2 examples and finalized the following 4 prominent patterns as codes:
\begin{itemize}
\setlength{\itemsep}{0.5pt}
  \item {\bf Refute}: Argue by directly giving a reason for why the partner's specific {\em answer} is believed to be incorrect. Examples: ``Small [partner choice] is the opposite of large [negative clue] and will not work''; ``Louisana isn't a country, therefore NonCountry applies.''
  \item {\bf Query}: Ask the partner to explain their answer, a part of their answer or ask for a clarification in their explanation. Examples: ``Why do you think it should be bill?''; ``How would bridge work?''
  \item {\bf Counter}: Pose a counter-argument to a partner {\em in response to} their explanation. Example: A: ``Erdogan's government is nationally affiliated with Turkey.'' B: ``[...] The sentence could be interpreted as one of Turkey's allies is helping them with the EU thing.''
  \item {\bf Previous}: Explicitly state that knowledge/line of reasoning acquired from a previous discussion is being used. Example: ``I had window at first too, but someone else had bridge, but they thought bridge because of the card game bridge, and that made sense to me'';
\end{itemize}

We found that workers used these contextual patterns frequently during their discussions for both domains with $77.6\%$ and $86.8\%$ of all discussions utilizing at least one pattern in the Relation Extraction and Codenames domains respectively. We can also see that distribution of patterns across the two domains (Table~\ref{tab:table-coding}) on discussions converging to the correct answer indicates that the utility of each pattern may be different in different domains. We hypothesize that the higher frequency of {\bf Counter} and lower frequency of {\bf Query} in Relation Extraction is likely due to the justification seeding which reduced need for workers to ask for explanations but encouraged more counter-arguments.

We also condensed the reasons for workers changing their answer down to 3 basic categories: learning about the {\em task} (rules), agreeing on meaning of concepts in a {\em question}, and being {\em convinced} by an argument. After coding the discussions, we found that the distribution of the reason for changing answers was $18\%, 3\%, 79\%$ for Relation Extraction domain and $17\%, 28\%, 55\%$ for Codenames, across each category (task, question, convinced) respectively showing that discussions could help workers understand the task.

\begin{table}
  \centering
  \begin{tabular}{|c|c|c|}
    \hline
    {} & Relation Extraction & Codenames \\
    \hline
    Refute     & {42\%} & {59\%} \\
    \hline
    Query      & {25\%} & {35\%} \\
    \hline
    Counter    & {34\%} & {14\%} \\
    \hline
    Previous   & {16\%} & {10\%} \\
    \hline
  \end{tabular}
  \caption{Proportion of each pattern appearing in discussions that converged to the correct answer for each domain.}
  \label{tab:table-coding}
  \vspace*{-0.15in}
\end{table}

We also observed that $70\%$ of all discussions and $75\%$ of discussions converging to the right answer used our rule shorthands when referring to the rules instead of describing them. However, we note that simply citing shorthands doesn't correlate with convergence of a discussion ($p > 0.1$).

\subsection{Do Workers Learn Through Discussion?}
While we didn't design discussions to be used as a way of training workers, many reported that they ``understood the task much better'' after discussions in pilot experiment feedback so we explored the effects of discussions on workers' future accuracy. We tested a worker's performance by adding post-test questions after they finished their corresponding experiment condition. We selected $4$ questions for the Relation Extraction domain and $1$ for the Codenames domain, all of comparable difficulty to the main questions, to be individually evaluated.

Average accuracy on the individual evaluation sections trended higher for the discussion condition: accuracies were $66.7\%$, $69.3\%$, and $73.9\%$ for the {\em baseline} (no argumentation), {\em reconsider} and {\em discussion} conditions respectively in the Relation Extraction domain and $46.7\%$ and $52.0\%$ for the {\em baseline} and {\em discussion} conditions in the Codenames domain. However, ANOVA on all conditions for each domain shows no statistically significant interaction ($F(1, 49.1) = 0.013, p > 0.1$ and $F(2, 58.3) = 1.03 , p > 0.1$ for Codenames and Relation Extraction respectively) between the experiment condition and the accuracy on the individual evaluation questions. We conjecture that need for argumentation may be reduced as workers better learn the guidelines through peer interaction~\cite{kulkarni2015peer,dow-cscw12}, but the difficult questions will likely always warrant some debate.

\section{Discussion}

\begin{figure}
  \centering
  \includegraphics[width=0.9\columnwidth]{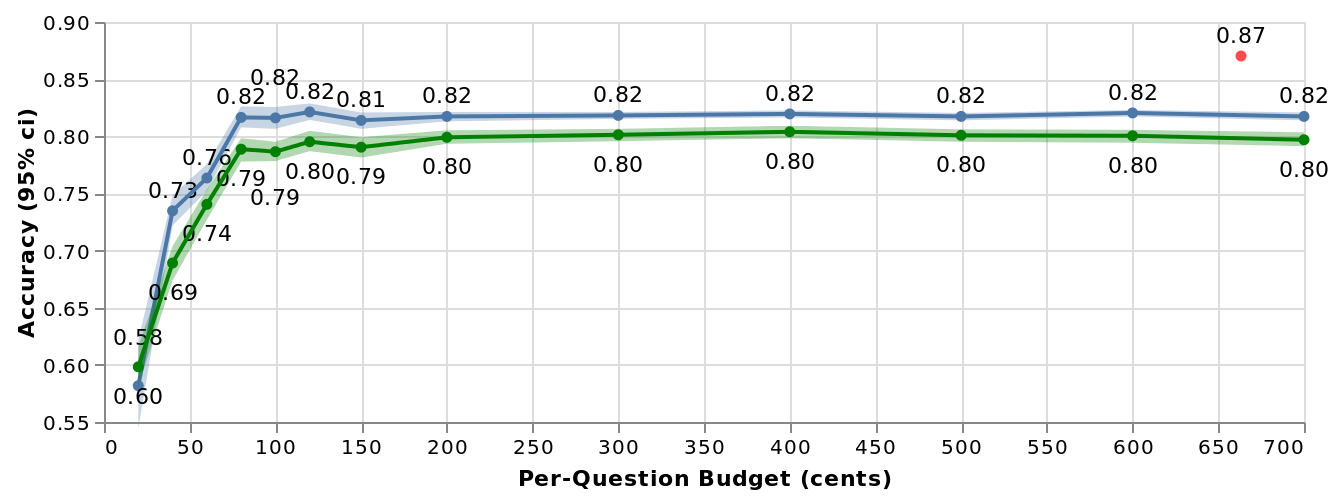}
  \caption{Scaling of majority vote (green) and EM-aggregated performance (blue) for one-shot argumentation (Microtalk) on the Relation Extraction domain, computed by simulation (100 simulations per budget) excluding training cost. Confirming previous reports~\protect\cite{demartini-www12,Drapeau2016MicroTalkUA},  we observe that accuracy plateaus. While expensive due to the use of real-time crowdsourcing, EM-aggregated performance of \implement\ (shown as a red dot) is higher.}
  \label{fig:plot-cost-comparison}
  \vspace{-0.15in}
\end{figure}

In this paper we explored the potential for multi-turn, contextual argumentation as a next step for improving crowdsourcing accuracy. Multi-turn argumentation systems provide some distinct benefits over one-shot reconsider systems. Since workers can respond directly to each other's arguments, it is much more likely that a correct, minority opinion may convince the majority without the need to recruit a huge number of workers. This power does come with caveats, noted earlier, such as higher costs and the need for  training workers how to recognize and make good arguments (in addition to training them on the base task). Our results suggest that workers can be trained to discuss successfully.

While each \textbf{discussion} task in \implement\ required more worker time, we found significantly higher gains to individual worker accuracy compared to the \textbf{reconsider} condition from {\sc MicroTalk}. We believe that much of the increase in work time stems from our decision to use synchronous, real-time crowdsourcing in \implement, leading to higher per-argument-task costs. Under a synchronous environment, workers must wait for other workers' actions during and in-between discussions. Since our experiments are focused on {\em evaluating} the multi-turn argumentation workflow, synchronized discussions allowed us to better collect data in a controlled way. Many efficiency optimizations, that we did not explore, could be implemented to run the \sys\ workflow at scale in a more cost effective way. Specifically, an asynchronous implementation of \sys\ would eliminate the need for workers to wait for each other, reducing costs. Alternatively, if a synchronous implementation were run at larger scale on a much larger set of problems, there would be proportionately less overhead. 

Argumentation (both contextual and one-shot) is likely unnecessary for many crowdsourcing tasks. For example, if one is merely labeling training data for supervised machine learning  (a common application), then it may be more cost effective to eschew most forms of quality control (majority vote, EM or argumentation) and instead collect a larger amount of noisy data~\cite{lin-hcomp14}. However, if one needs data of the highest possible accuracy, then argumentation --- specifically contextual, multi-turn argumentation --- is the best option. \sys\ may cost more than other approaches, but as shown in Figure~\ref{fig:plot-cost-comparison}, it achieves higher accuracy than any other approach, regardless of cost.
Furthermore, the \sys\ workflow can handle questions with many possible answers such as those in Codenames. Single-shot argumentation systems, such as {\sc MicroTalk}, aren't practical in these situations, because of the need to pre-collect arguments against so many alternative answers.  

In the end, the most cost effective crowd technique depends on both problem difficulty and quality requirements. 

High cost methods, like argumentation, should be reserved for the most difficult tasks, such as developing challenging machine learning {\em test} sets, or tasks comprising a  high-stakes decision, where a corresponding explanation is desirable.

\section{Conclusion \&\ Future Work}

In this paper, we presented \sys, a novel workflow that engages workers in {\em multi-turn, contextual} argumentation (discussion) to improve crowd accuracy on difficult tasks. We implemented this workflow using a synchronous, real-time design for discussions tasks and created the \implement\ system. Since the quality of a discussion depends on its participants, we also designed and implemented gated instructions and a novel justification training task for \implement\ to ensure competent discussions through improving workers' ability to recognize and synthesize good arguments.

We demonstrate that our implementation of \implement, the synchronous version of the \sys\ workflow, is able to achieve
\begin{itemize}
\setlength{\itemsep}{0.5pt}
\setlength{\parsep}{1pt}
\item Higher improvement of individual worker accuracy compared to a state-of-art, one-shot argumentation system on a difficult NLP annotation task: 16.7 {\em vs.} 6.8 percentage points improvement, at a higher cost, and
\item Very high accuracy in a non-binary choice answer task that would be impractical with one-shot argumentation: 98.8\% accuracy (a 32.1 percentage point improvement over the initial accuracy)
\end{itemize}

Both these accuracies are much higher than can be achieved without argumentation. Traditional majority vote and expectation-maximization without argumentation approaches plateau at 65\% on similar questions~\cite{Drapeau2016MicroTalkUA}. Additionally, we observed several interesting patterns of discourse that are enabled by multi-turn, contextual argumentation and note that many successful discussions utilize these patterns.

There are many future directions for improving the argumentation workflow and system implementation. Currently, the cost of argumentation is still relatively high but cost may be reduced further. It's common for people to context switch between several ongoing discussions as well as have group discussions with many people, both of which could be interesting modifications to the workflow. We also envision that better models of discussions could allow a system to only pair arguments where the outcome reduces uncertainty.

Furthermore, there is potential in incorporating natural language processing techniques to identify and support positive behavior patterns during argumentation and opportunities for learning from misconceptions surfaced during discussion to improve training and task instructions. 

Finally, we hope to apply argumentation techniques to a wider range of tasks and meta-tasks, including issues studied in Turkomatic~\cite{kulkarni-cscw12} and flash teams~\cite{retelny-uist14} --- the process of defining a problem and refining a workflow to achieve it.
\section{Acknowledgements}
We would like to thank Eunice Jun, Gagan Bansal and Tongshuang Wu for their helpful feedback and participation in pilot experiments. This work was supported in part by NSF grant IIS-1420667, ONR grants N00014-15-1-2774 and N00014-18-1-2193, the WRF/Cable Professorship and support from Google.


\balance{}


\bibliographystyle{SIGCHI-Reference-Format}
\bibliography{main,general}

\end{document}